\documentclass[12pt]{article} \usepackage{dina4p} \usepackage{my}
\usepackage{epsfig,amsmath} \usepackage{graphics}

\newcommand{\CCFM}{CCFMa,CCFMb,CCFMc,CCFMd}
\newcommand{\BFKL}{BFKLa,BFKLb,BFKLc}

\newcommand{\PYTHIAMC}{Pythia61}

\def\JETSET {{\sc Jetset}/{\sc Pythia}}

\newcommand{\CASCADEMC}{CASCADE,jung_salam_2000,CASCADEMC}
\newcommand{\DGLAP}{DGLAPa,DGLAPb,DGLAPc,DGLAPd}

\newcommand{\as}{\alpha_\mathrm{s}}

\def\CASCADE{{\sc Cascade}}

\def\J {J/\psi}

\def\cc {$c\bar{c}$~}

\def\lsim{\mathrel{\rlap{\lower4pt\hbox{\hskip1pt$\sim$}}
    \raise1pt\hbox{$<$}}}                
\def\gsim{\mathrel{\rlap{\lower4pt\hbox{\hskip1pt$\sim$}}
    \raise1pt\hbox{$>$}}}                
 
\renewcommand{\thefootnote}{\fnsymbol{footnote}} 
\begin{document} 
\begin{flushright}
DESY 02-017\\
LUNFD6/(NFFL--7209) 2002 \\
hep-ph/0203025\\
\end{flushright}
\vspace*{10mm}
\begin{center}  \begin{Large} \begin{bf}
  A phenomenological interpretation of open charm production \\  
  at HERA in terms of the semi-hard approach \\ 
  \end{bf}  \end{Large}
  \vspace*{5mm}
  \renewcommand{\thefootnote}{\arabic{footnote}}
  \begin{large}
S.P.~Baranov  
\footnote[1]{\parbox[t]{10cm}{E-mail: {\tt baranov@sci.lebedev.ru}}}\\ 
{\sl Lebedev~Institute~of~Physics, 
Leninsky~prosp.~53, Moscow~117924, Russia} \\
H.~Jung   
\footnote[2]{\parbox[t]{10cm}{E-mail: {\tt jung@mail.desy.de}}}
L. J\"onsson 
\footnote[3]{\parbox[t]{10cm}{E-mail: {\tt leif@quark.lu.se}}}\\  
{\sl Department of Elementary~Particle~Physics, Lund~University, 
22100~Lund,~Sweden} \\ 
S. Padhi \footnote[4]{\parbox[t]{10cm}{E-mail: {\tt Sanjay.Padhi@desy.de}}}\\ 
{\sl Department~of~Physics, McGill~University, Montreal,
 Quebec, Canada, H3A~2T8  } \\ 
N.P.~Zotov 
\footnote[5]{\parbox[t]{10cm}{E-mail: {\tt zotov@theory.sinp.msu.ru}}}\\ 
{\sl Skobeltsyn~Institute~of~Nuclear~Physics,  
Moscow~State~University, Moscow~119992,~Russia} 
  \end{large}%
\end{center}
\begin{quotation}
\noindent
{\bf Abstract:}  In the framework of the 
semi-hard ($k_t$-factorization) approach, we analyze the 
 various charm production processes in the kinematic region covered by
 the HERA experiments.
\end{quotation}
 \renewcommand{\thefootnote}{\arabic{footnote}} 
 \setcounter{footnote}{0} 
 
\section{Introduction}
 
 At the energies of modern 
lepton-hadron and hadron-hadron colliders,
the interaction dynamics is governed 
 by the properties of parton distributions in the small $x$ region. 
This domain is characterized by the double inequality $s\gg\mu^2\gg\Lambda^2$, 
which shows that the typical parton interaction scale $\mu$ is much higher 
than the QCD parameter $\Lambda$, but is much lower than the total c.m.s. 
energy $\sqrt s$. The situation is therefore classified as ``semi-hard''. 
In such a case, the perturbative QCD expansions in $\alpha_s$ may contain 
large coefficients ${\cal O}\left[\log(s/\mu^2)\right]=
{\cal O}\left[\log(1/x)\right]$ which 
compensate the smallness of the coupling constant $\alpha_s(\mu^2/\Lambda^2)$. 
The resummation \cite{Gribov,Collins} of the terms 
$\left[\log(1/x)\,\alpha_s\right]^n$ results in the so called unintegrated 
parton distribution 
${\cal F}(x,k_{t}^2)$, which determines the probability to find a parton 
carrying the longitudinal momentum fraction $x$ and transverse 
momentum $k_t$. 
If the terms $\left[\log(\mu^2/\Lambda^2)\,\alpha_s \right]^n$ and
$\left[\log(\mu^2/\Lambda^2)\,\log(1/x)\,\alpha_s\right]^n$   
are also
resummed, then the unintegrated parton distribution depends also
on the probing scale $\mu$, and will be labeled as  
${\cal A}(x,k_{t}^2,\mu^2)$. 
That generalizes the factorization of the 
hadronic matrix elements beyond the collinear approximation (hereafter 
this generalized factorization will be referred to as 
``$k_t$ factorization''~\cite{Collins,CCH}). The  unintegrated
parton distributions obey certain evolution equations  
(e.g., BFKL \cite{\BFKL} or CCFM \cite{\CCFM}) and are related to the  
conventional DGLAP \cite{\DGLAP} densities once the $k_{t}$  
dependence is integrated out.  
Nowadays, the significance of the $k_t$ factorization (semi-hard) approach 
 becomes more and more 
commonly recognized. Its applications to a variety of photo-, lepto- and 
hadro-production processes are widely discussed in the literature  
\cite{Shaba,Shabb,Shabc,ZotLipa,ZotLipb,d*gam,d*dis1,d*dis2,%
CASCADE,jung_salam_2000,xgam,Maria,Hagler_bbar,jung-hq-2001,Teryaev_jpsi2,%
Teryaev_jpsi1,chao_jpsi_1,chao_jpsi_2}. 
Remarkable agreement is 
found between the data and the theoretical calculations regarding  
photo- \cite{d*gam} and electro-production \cite{d*dis1,d*dis2} of $D^*$ 
mesons, and of forward jets \cite{CASCADE,jung_salam_2000},   
as well as for specific kinematic correlations observed in 
photoproduction of $D^*$ mesons associated with jets~\cite{xgam} at HERA. 
Also in hadro-production of beauty \cite{Maria,Hagler_bbar,jung-hq-2001}, 
$\chi_c$ \cite{Teryaev_jpsi2} 
and $\J$~ \cite{Teryaev_jpsi1,chao_jpsi_1,chao_jpsi_2} at the TEVATRON  
good agreement is observed.    
However, for a consistent application of $k_t$-factorization 
in different models, the unintegrated
gluon distribution has to be determined  in the same framework.
Also the various approximations needed to describe the
experimental data have to be carefully investigated.
In the present paper we have attempted a systematic comparison 
of model predictions with experimental data
regarding the heavy flavor production processes 
at HERA. 
\section{The \boldmath$k_t$-factorization approach applied to charm production}
The production of
 open-flavored \cc pairs in $ep$-collisions 
is described in terms of the photon-gluon fusion mechanism. 
A generalization of the usual parton model to the $k_t$-factorization 
approach implies two essential steps. These are the introduction of 
unintegrated gluon distributions and the modification of the gluon spin 
density matrix in the parton-level matrix elements.  
\par
Here we consider only {$\gamma^* g^* \to c \bar{c}$}.
Let $k_{\gamma}$, $k_g$, $p_c$ and $p_{\bar c}$ be the four-momenta of the 
initial state photon,
the initial state gluon, the final state quark and anti-quark 
respectively, and $\epsilon_{\gamma}$ and $\epsilon_g$ are
the corresponding polarization vectors.
The photon-gluon fusion matrix elements  for the production
of an open-flavored $c\bar c$ pair then reads (with a charm mass $m_c$):
\begin{eqnarray}\label{Md}
{\cal M}(\gamma g\to c\bar{c})& = &\bar{u}(p_c)\left(
  \frac{
  \not\epsilon_{\gamma}\,(\not p_c -\not k_{\gamma}+m_c)\not\epsilon_{g}}   
  {k_{\gamma}^2-2k_{\gamma}p_c} 
 + 
\frac{\not\epsilon_{g}\,(\not p_c -\not k_g+m_c)\not\epsilon_{\gamma}}
  {k_g^2-2k_gp_c} \right)u(p_{\bar{c}})
\end{eqnarray}
The matrix-element squared for open heavy quark production has already been
calculated in~\cite{Collins,CCH}, which we label CE-CCH in the following. 
In \cite{ZotSala} (labeled as SZ) the calculation of the matrix 
elements for open heavy quark production has been repeated.
In \cite{d*gam,xgam} (labeled as BZ) 
the method of orthogonal amplitudes~\cite{prange} was applied.
When calculating the spin average of the matrix element squared, 
BZ uses $L^{\mu\nu}$ for the 
photon polarization matrix: 
\begin{equation} \label{epsgam} 
L^{\mu\nu} = \overline{\epsilon_{\gamma}^{\mu}\epsilon_{\gamma}^{*\nu}}= 
4\pi\alpha \left[8p_e^{\mu}p_e^{\nu}-4(p_ek_{\gamma})g^{\mu\nu}\right]
/(k_{\gamma}^2)^2 
\end{equation}
where $p_e$ is the four momentum of the incoming electron. The expression 
also includes the photon propagator factor and photon-lepton coupling. 
In the calculation of CE-CCH and SZ
the photon is treated in a similar way 
as was the gluon in~\cite{Collins}:
\begin{equation} \label{epsglu} 
G^{\mu\nu} = \overline{\epsilon_g^{\mu}\epsilon_g^{*\nu}}= 
k_{t\;g}^{\mu} k_{t\;g}^{\nu}/|k_{t\;g}|^2. 
\end{equation} 
This formula converges to the usual 
$\sum\epsilon^{\mu}\epsilon^{*\nu}=-g^{\mu\nu}$ when $k_{t}\to 0$. 
\par 
In BZ the complete set of   
matrix elements have been tested for gauge invariance by substituting the gluon  
momenta with their polarization vectors showing  explicitly
the gauge invariance of the matrix element in order ${\cal O}(\alpha_s)$.  
\par
The hard scattering cross section for a boson gluon fusion process is written 
as
a convolution of the partonic cross section 
$\hat{\sigma}(x_g,k_{t};\; {\gamma^* g^* \to q \bar{q}})$ with
the $k_{t}$ dependent (unintegrated) gluon density ${\cal A}(x,k_t^2,\mu^2)$
(here and in the following $k_{t}\;(k_{t\;\gamma})$ is a shorthand notation for 
 $|\vec{k}_{t}|$ ($|\vec{k}_{t\;\gamma}|$)
  with $\vec{k}_{t}$ ($\vec{k}_{t\;\gamma}$)
  being the two-dimensional vector of the
 transverse momentum of the gluon (photon)):
\begin{equation}
\sigma = \int dk_{t}^2 dx_g {\cal A}(x_g,k_{t}^2,\mu^2)
 \hat{\sigma}(x_g,k_{t};\;{\gamma^* g^* \to q \bar{q}})\,,
\label{x_section}
\end{equation}
with the off-shell matrix elements either given by 
CE-CCH, SZ or by BZ. 
The multidimensional integrations can be 
performed by means of Monte-Carlo technique either by using    
VEGAS~\cite{VEGAS} for the pure parton level calculations,
or by using the full Monte Carlo event generator  \CASCADE ~\cite{\CASCADEMC}.
\begin{table}[htb]
\begin{center}
\begin{tabular}{|c|c|c|c|}
\hline
  & CE-CCH & SZ & 
  BZ
  \\ \hline  
$k_{t\;\gamma} = k_{t} =0$~GeV &  1
                                   &  1
					     &  0.9
 \\ \hline  
$k_{t\;\gamma} = 0$~GeV, $k_{t} =10$~GeV & 1
                                            & 1
                                            & 0.86
 \\ \hline  
$k_{t\;\gamma} = 10$~GeV, $k_{t} =0$~GeV  & 1
                                                 & 1
                                                 & 0.96
 \\ \hline  
$k_{t\;\gamma} = 10$~GeV, $k_{t} =10$~GeV  & 1
                                                  & 1
                                                  & 0.93
 \\ \hline  
\end{tabular}
\caption{\it  
 Comparison of different calculation of the matrix elements for 
$\gamma^* g^* \to c \bar{c} $. Shown are the matrix elements normalized to the
matrix element CE-CCH for
$p_{t\;c} =  5$~GeV, $\eta_c = \eta_{\bar{c}}=0$ in the $ep$ c.m.s. with $\sqrt{s}=300$~GeV. For
the comparisons the momenta of the incoming and outgoing partons have 
been modified to satisfy the small $x$ requirement: 
$k_{\gamma} = x_{\gamma} p_e + \vec{k}_{t\;\gamma}$ and 
$k_{g} = x_{g} p_p + \vec{k}_{t}$  resulting in 
$k_{\gamma}^2 = -k_{t\;\gamma}^2$ and $k_{g}^2 = -k_{t}^2$.
}
\label{xsect}
\end{center}
\end{table}
Since it is difficult to compare the different matrix elements with each other
analytically and to prove that they agree, we have performed several numerical
checks. 
In Tab.~\ref{xsect} we show a numerical comparison of the different matrix
elements for $\gamma^* g^* \to c \bar{c} $. For the comparison we have chosen
the transverse momentum of the charm quark to be
$p_{t\;c} =  5$~GeV, its rapidity to be
$\eta_c =\eta_{\bar{c}} = 0$ in the $ep$ c.m.s. 
and calculated the other kinematic quantities accordingly. The transverse
momenta of the incoming partons are as indicated in the table.
In addition we have removed all $\alpha_s$
dependencies from the matrix elements. In the calculation of the off-shell
matrix elements, approximations are necessary to satisfy the $k_t$-factorization
theorem: the gluon polarization tensor as given in eq.(\ref{epsglu})
is applied (which is different to the full polarization tensor of
eq.(\ref{epsgam})),
and the transverse momentum must be
dominating the virtuality: $k^2 = - k_t^2$. The last
condition is essentially the small $x$ (or high energy) approximation.
These criteria are satisfied in the calculations of 
CE-CCH and SZ, whereas
the calculation of BZ is done using the full
polarization tensor $L^{\mu\nu}$ for the photon and only $G^{\mu\nu}$ 
for the gluon without
applying the small $x$ approximation to the gluon four-vector. 
To study the effect of the
different approximations, 
we have compared the matrix elements at large c.m.s
energies of $\sqrt{s}=30000$~GeV and observe good agreement for all cases. This
means, that in the asymptotic limit of very high energies, the small $x$
approximation and the use of  $G^{\mu\nu}$ are justified.
In Tab.~\ref{xsect} we compare the  matrix elements at a c.m.s
energy of $\sqrt{s}=300$~GeV, typical for HERA  experiments, 
for different values of the transverse photon momentum
$k_{t\;\gamma}$ and the incoming gluon $k_{t}$.
Whereas the calculations of 
CE-CCH and SZ agree
perfectly in all cases, a systematic difference of the order of $\sim$ 10\% 
to the calculation of BZ is observed. Since we
obtained agreement in the high energy limit, this can be attributed to the
effect of the small $x$ approximation. 
In addition BZ treat the polarization of
the photon and the gluon differently, and therefore this effect can be 
quantified by observing a difference of $\sim$ 10\% between the rows two and 
three in Tab.~\ref{xsect}. The above investigations indicate that 
the effects of the small $x$ approximation applied
at HERA energies are of the order of $\sim$ 10\%.
\par
It is also interesting to consider the limit $k_{t} \to 0$ of the matrix
elements. To do that we define a reduced cross section $\tilde{\sigma}$:
\begin{equation}
\tilde{\sigma}(k_{t}) = \int d\mbox{Lips} \;|ME|^2 
\end{equation}
where we integrate over the Lorentz-invariant-phase-space ($\mbox{Lips}$) of the
final state quarks. The matrix element $|ME|^2 $ is taken from 
CE-CCH,
where  we have set $16 \pi^2 \alpha_{em} \alpha_s e_q^2 \equiv 1 $.
\begin{figure}[htb]
\begin{center} 
  \vspace*{1mm} 
\includegraphics[width=0.48\linewidth]{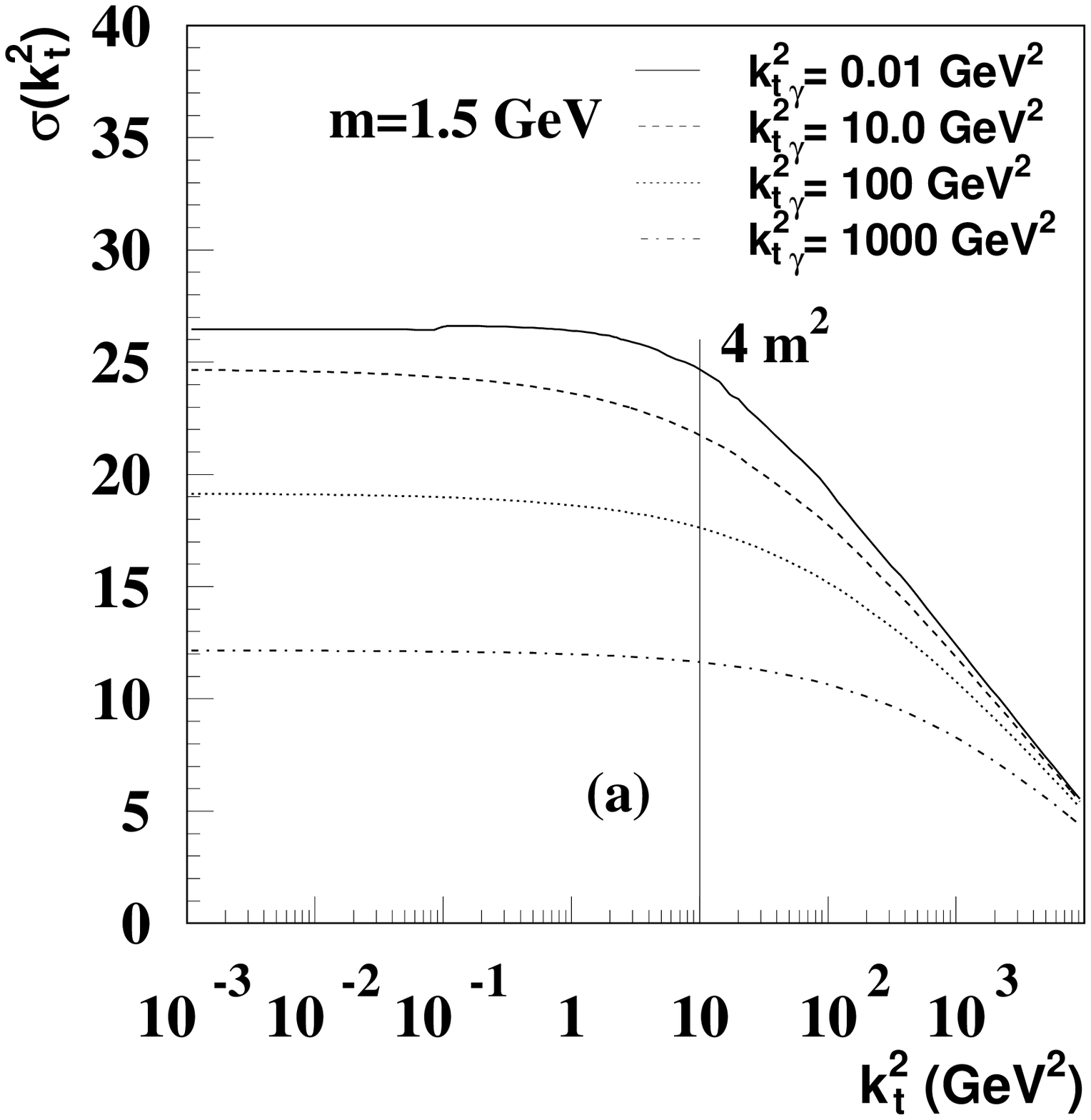}
\includegraphics[width=0.48\linewidth]{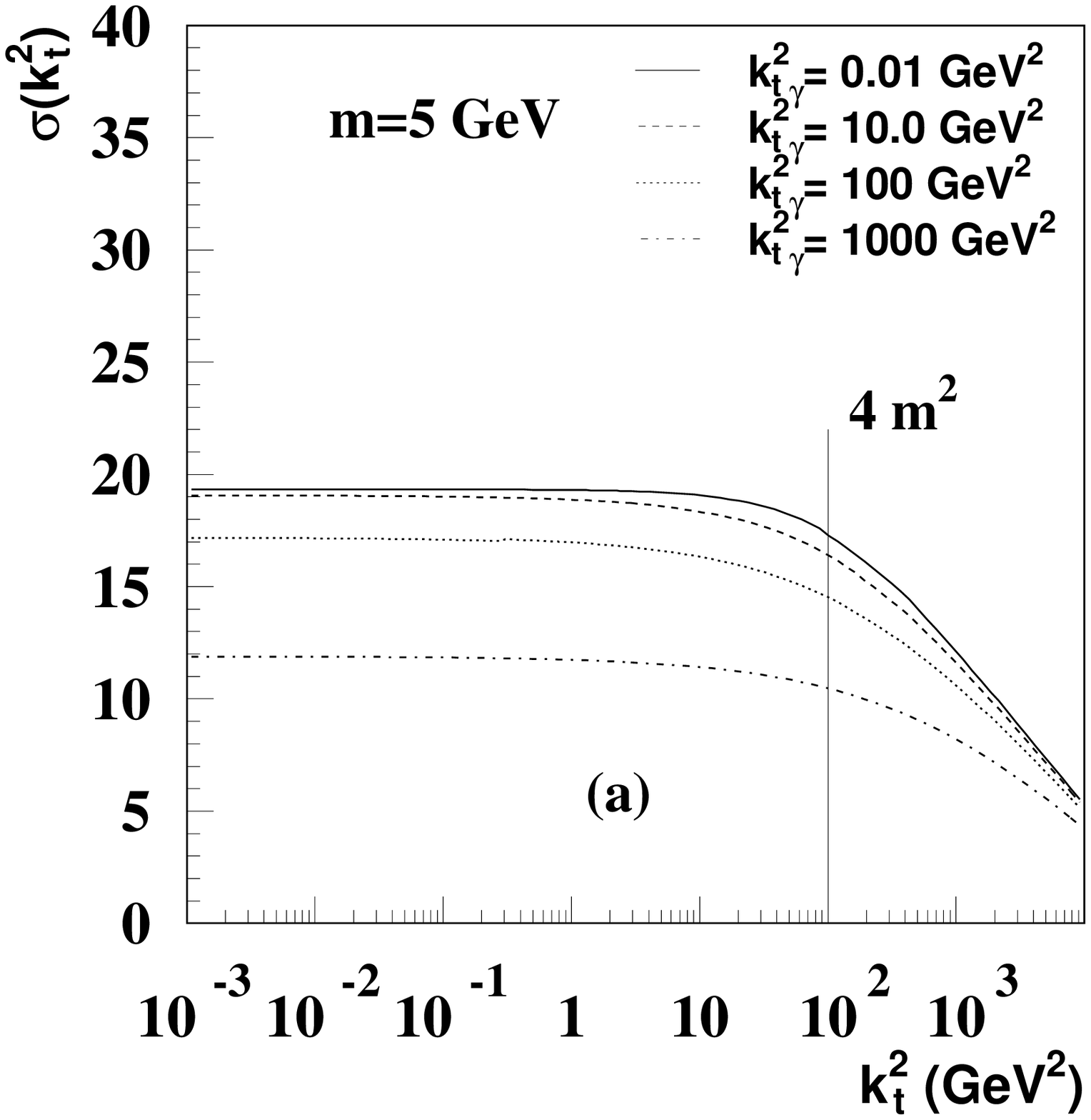}
\caption{{\it 
The reduced cross section $\tilde{\sigma}(k_{t})$ 
as a function of the transverse
momentum $k_{t}$ of the incoming gluon for different values of the transverse
momentum of the incoming photon $k_{t\;\gamma}$  
($m=1.5$~GeV in $(a)$, $m=5$~GeV in $(b)$, 
 $\sqrt{s}=30000$~GeV and a fixed $x_{\gamma}=x_g=0.01$).
 }}\label{ME-kt} 
\end{center}
\end{figure} 
In Fig.~\ref{ME-kt} we show 
$\tilde{\sigma}(k_{t})$ as a function
of the transverse momentum of the incoming gluon $k_{t}$ for 
quark masses of $m=1.5$~GeV in 
Fig.~\ref{ME-kt}$a$ and for $m=5$~GeV in Fig.~\ref{ME-kt}$b$ using
$\sqrt{s}=30000$~GeV and a fixed $x_{\gamma}=x_g=0.01$. In both cases 
a smooth behavior for $k_{t} \to 0$ is observed. 
It is also interesting to
note that in all cases
the cross section starts to
decrease at $k_{t}^2 \gsim 4 m^2$.
The region $k_{t}^2 > 4 m^2$ is still contributing to the
total cross section significantly, showing one of the main differences to 
the usual collinear approximation, where this region is completely ignored. 
\section{The unintegrated gluon distributions}
Cross section calculations require an explicit representation of the 
$k_t$ dependent (unintegrated) gluon density ${\cal A}(x,k_t^2,\mu^2)$. 
We have used three different representations, one ({\it JB}) coming from a 
leading-order perturbative solution of the BFKL 
equations~\cite{Bluem}, the second set ({\it JS})
derived from a numerical solution of the CCFM
equation~\cite{CASCADE,jung_salam_2000} and the third ({\it KMR})
from solution of a
combination of the BFKL and DGLAP equations~\cite{martin_kimber}.
\begin{defl}{123456789}
  \item[{\boldmath$JB$}] 
The unintegrated gluon density
 ${\cal A}(x,k_{t}^2,\mu^2)$, in the approach of~\cite{Bluem}, 
is calculated as a convolution of the ordinary 
gluon density $xG(x,\mu^2)$ (here we use GRV~\cite{GRV95})
with universal weight factors: 
\begin{equation} \label{conv} 
 {\cal A} (x,k_{t}^2,\mu^2) = \int_x^1 
 {\cal G}(z ,k_{t}^2,\mu^2)\, 
 \frac{x}{z }\,G(\frac{x}{z },\mu^2)\,dz , 
\end{equation} 
\begin{equation} \label{J0} 
 {\cal G}(z ,k_{t}^2,\mu^2)=\frac{\bar{\alpha}_s}{z \,k_{t}^2}\, 
 J_0(2\sqrt{\bar{\alpha}_s\ln(1/z )\ln(\mu^2/k_{t}^2)}), 
 \qquad k_{t}^2<\mu^2, 
\end{equation} 
\begin{equation}\label{I0} 
 {\cal G}(z ,k_{t}^2,\mu^2)=\frac{\bar{\alpha}_s}{z \,k_{t}^2}\, 
 I_0(2\sqrt{\bar{\alpha}_s\ln(1/z )\ln(k_{t}^2/\mu^2)}), 
 \qquad k_{t}^2>\mu^2, 
\end{equation} 
where $J_0$ and $I_0$ stand for Bessel functions (of real and imaginary 
arguments, respectively), and 
$\bar{\alpha}_s=3 \as/\pi$ is connected to the pomeron intercept 
$\alpha(0) = 1 + \Delta$, with 
$\Delta=\bar{\alpha}_s 4 \log 2 \label{Delta_bfkl}$ in LO.
An expression for 
$\Delta$ in NLO is given in~\cite{NLLFL}: 
 $\Delta=\bar{\alpha}_s 4 \log 2- N\bar{\alpha}_s^2$.  
In our calculations presented here we use 
the solution of the LO BFKL equation and  
treat $\Delta$ as free parameter varying between $0.166 < \Delta < 0.53$ with a
central value of $\Delta = 0.35$. 
\item[{\boldmath$ JS$}]
  The CCFM evolution equations have been solved 
  numerically in~\cite{CASCADE,jung_salam_2000} using a Monte Carlo method. 
  According to the
  CCFM evolution equation, the emission of partons during the initial
  cascade is only allowed in an angular-ordered region of phase space.
  The maximum allowed angle $\Xi $ for any gluon emission sets the
  scale $\mu^2$ for the evolution and is defined by the hard
  scattering quark box, which connects the exchanged gluon to the
  virtual photon. 
  \par
  The free parameters of the starting gluon distribution were fitted to
  the structure function $F_2(x,Q^2)$ in the range $x < 10^{-2}$ and 
  $Q^2 > 5$~GeV$^2$ as described
  in~\cite{jung_salam_2000}. 
\item[{\boldmath$ KMR$}]
  In {\it KMR}~\cite{martin_kimber} the dependence of the unintegrated gluon
  distribution on the two scales $k_t^2$ and $\mu^2$ was investigated: the scale
  $\mu^2$ plays a dual role, it acts as the factorization scale and also controls
  the angular ordering of the partons emitted in the evolution.
  This results in a form  similar to the 
  differential form of the CCFM
  equation, however 
  the splitting function $P(z)$  is 
  taken from the single scale evolution of the {\it unified} DGLAP-BFKL
  expression discussed in \cite{Martin_Stasto}.
  The unintegrated gluon density
  $x{\cal A}(x,k_t^2,\mu^2)$ covering the whole range in $k_t^2$ has
  been evaluated by~\cite{Kimber_pc_2001}, giving:
  \begin{equation}
  x{\cal A}(x,k_t^2,\mu^2) = \left\{\begin{array}{ll}
        \frac{xG(x,k_{t0}^2)}{k_{t0}^2} & \mbox{if  } k_t < k_{t0} \\
     \frac{f(x,k_t^2,\mu^2)}{k_t^2} & \mbox{if  } k_t \geq k_{t0}   
  \end{array} \right.
  \end{equation}
  with $xG(x,k_{t0}^2)$ being the integrated MRST~\cite{MRST} 
  gluon density function
  and $f(x,k_t^2,\mu^2)$ being the unintegrated gluon density 
  of~\cite{martin_kimber} starting from $k_t^2 > k_{t0}^2=1$~GeV$^2$. The
  unintegrated gluon density $x{\cal A}(x,k_t^2,\mu^2)$ therefore is
  normalized to the MRST function when integrated up to $k_{t0}^2$. 
\end{defl}
\par
\begin{figure}[htb] 
  \vspace*{1mm} 
  \begin{center}
\epsfig{figure=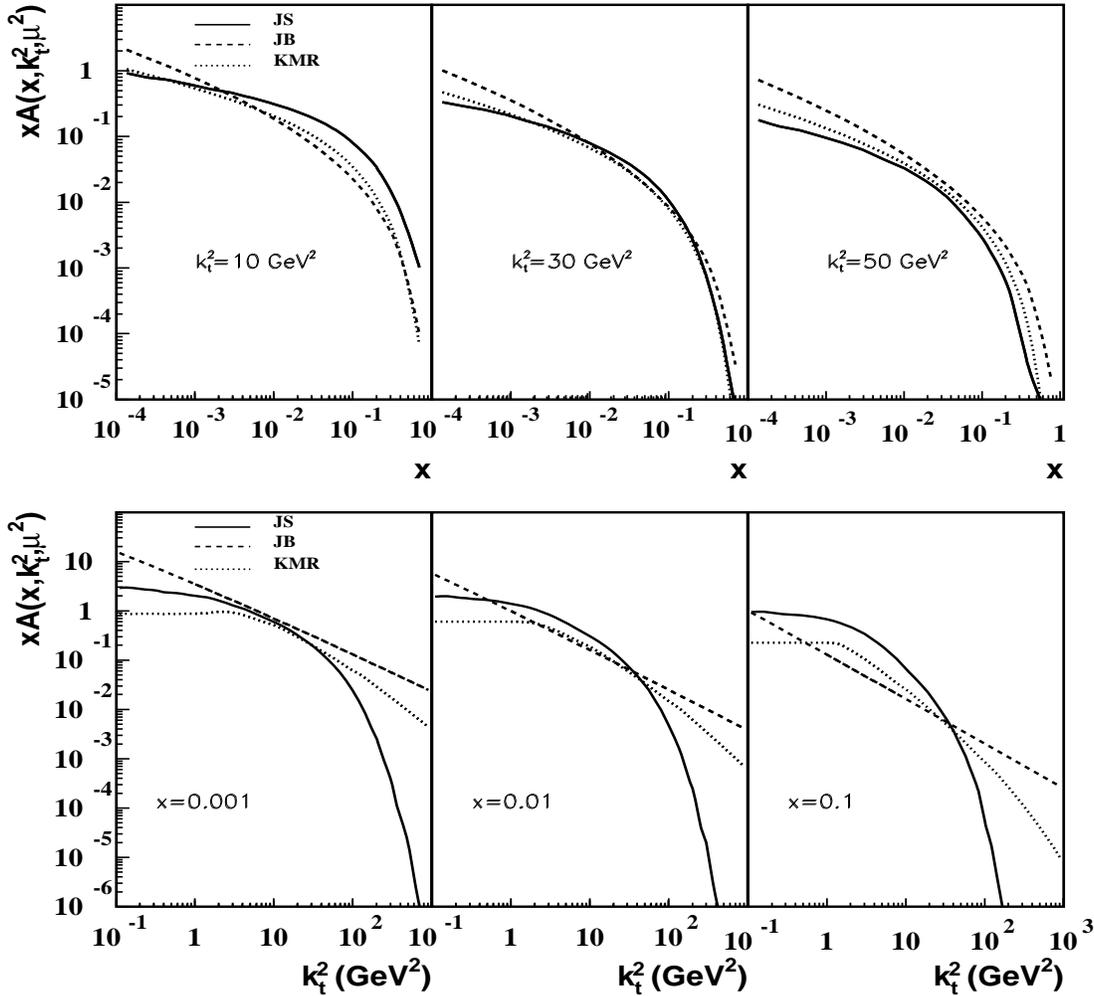, 
width=17cm,height=15cm} 
\caption{{\it 
The  $k_t$ dependent (unintegrated) gluon density at $\mu^2 =100$ GeV$^2$ 
as a function of 
$x$ for different values of $k_t^2$ (upper part) and as a function of 
$k_t^2$ for different values of $x$ (lower part) as given by  
{\it JS}~\protect\cite{CASCADE,jung_salam_2000} (solid line) , 
{\it JB}~\protect\cite{Bluem} (dashed line) and
{\it KMR}~\protect\cite{martin_kimber,Kimber_pc_2001} (dotted line). 
 }}\label{gluon} 
 \end{center}
\end{figure}

In Fig.~\ref{gluon} we show a comparison of the gluon density distributions  
at $\mu^2=100$ GeV$^2$ obtained from {\it JB}, 
 {\it JS} and {\it KMR} as a function
of $x$ for  different values of $k_t^2$  and
as a function of $k_t^2$ for different values of $x$. 
\par
From Fig.~\ref{gluon} we see that all three 
unintegrated gluon distributions 
show a significantly different behavior as a function of $x$ but even more as a
function of $k_t$. It will be interesting to see, how this different behavior
is reflected in the prediction of cross sections for experimentally observable
quantities like charm production at HERA.
\section{Numerical results and discussion}  
A comparison between model predictions and data in principle has to be made on 
hadron level and only if it turns out that hadronization effects are small  
will a comparison to parton level predictions make sense. 
However, a full simulation even of the partonic final state, including the 
initial and final state QCD cascade needs a full  Monte Carlo event
generator. Such a Monte Carlo generator based on $k_t$-factorization
and using explicitly off-shell matrix elements for the hard scattering process
convoluted with $k_t$-unintegrated gluon densities 
is presently only offered by the \CASCADE ~\cite{\CASCADEMC}
program which uses the CCFM unintegrated gluon distribution. This is because
only the CCFM evolution equation gives a description on how to explicitly build
the initial state gluon radiation by applying angular ordering.
Other sets of unintegrated gluon distribution can only be used to 
calculate quantities at the matrix element level, which can be compared to data 
only if the effect of hadronization and of the complete initial state parton  
cascade are insignificant. 
\par
In the following we want to systematically compare the predictions from the
$k_t$-factorization approach to  
published data on charm production at HERA.
For this we use $D^*$ photo-production data from ZEUS \cite{ZEUS-D*-gammap} and 
$D^*$ production in deep inelastic scattering from both 
ZEUS~\cite{ZEUS-D*-dis} and H1~\cite{H1-D*-dis}.
\begin{table}[htb]
\begin{center}
\begin{tabular}{|c|c|c|c|c|}
\hline
 ZEUS ($\gamma p$)~\protect\cite{ZEUS-D*-gammap} 
 & $130 <  W < 280$~GeV &$Q^2 < 1$~GeV$^2$ & $|\eta |< 1.5$
  & $p_t > 2$ GeV \\ \hline  
 ZEUS DIS~\protect\cite{ZEUS-D*-dis}
 &  $ 0.02 < y < 0.7 $ & $ 1 <Q^2 < 600 $~GeV$^2$ & 
$|\eta |< 1.5$ & $1.5 < p_t < 15$~GeV  \\ \hline  
H1 DIS~\protect\cite{H1-D*-dis}
 &  $ 0.05 < y < 0.7 $ & $ 1 <Q^2 < 100 $~GeV$^2$ &$|\eta |< 1.5$ & 
$1.5 < p_t < 15$~GeV\\ \hline 
\end{tabular}
\caption{\it 
Kinematic range of the different data used for comparison}
\end{center}
\end{table}
To do this we first calculate observables using a pure parton level calculation
based on the matrix element calculation of BZ
including the Peterson fragmentation
function~\cite{Peterson}
 for the transition from the charm quark to the observed $D^*$ meson,
where the $D^*$ meson is assumed to take a momentum fraction $z$ of the 
charm quark, in the $c\bar{c}$ center-of-mass (c.m.s.) frame. 
Then we compare the result with a full hadron level simulation using the Monte
Carlo generator \CASCADE\  with the matrix element of 
CE-CCH.
 Also here the Peterson fragmentation function 
is used but now
with $z$ being defined as the light-cone momentum fraction in the 
center-of-mass system 
of the string connecting the charm quark with its light quark partner, 
as implemented in \JETSET ~\cite{\PYTHIAMC}. 
We choose the {\it JS} unintegrated gluon for this comparison, which is 
also appropriate for a description of heavy quark production at high 
energies~\cite{jung-hq-2001}. 
\par
Next we investigate on parton level different unintegrated gluon densities.
After the optimal choice of model parameters has been found for the 
{\it JB} gluon density,  
giving the best possible agreement with data, 
we show a comparison to the {\it JS} and {\it KMR} unintegrated gluon density.
We then study the sensitivity of the model predictions to the details 
of the unintegrated gluon density, 
the charm mass and the scale.
\par
We also consider the rapidity distribution
of the produced $D^*$, which is very sensitive to the choice
of the unintegrated gluon density and the details of the
$c \to D^*$ fragmentation.
\par
Then we investigate the $x_{\gamma}$ distribution,
which is sensitive to the details of the initial state cascade.
We compare the predictions from a pure parton level calculation and a full event
simulation of \CASCADE\ with the measurements.
\par
At the end we show, motivated by preliminary studies
of ZEUS~\cite{ZEUS_eps499}, predictions which are
sensitive to the details of the heavy quark production mechanism.
\subsection{Transverse momentum distribution of {\boldmath$D^*$} mesons: 
comparison of parton and hadron level}
One observable  which is expected to show only little sensitivity to  
the hadronization and to the full 
simulation of the initial and final state QCD cascades is the transverse 
momentum $p_t$ of the $D^*$ meson in photo-production and deep inelastic 
scattering.  
\begin{figure}[htb]
\vskip -0.5cm
\begin{center}
\epsfig{figure=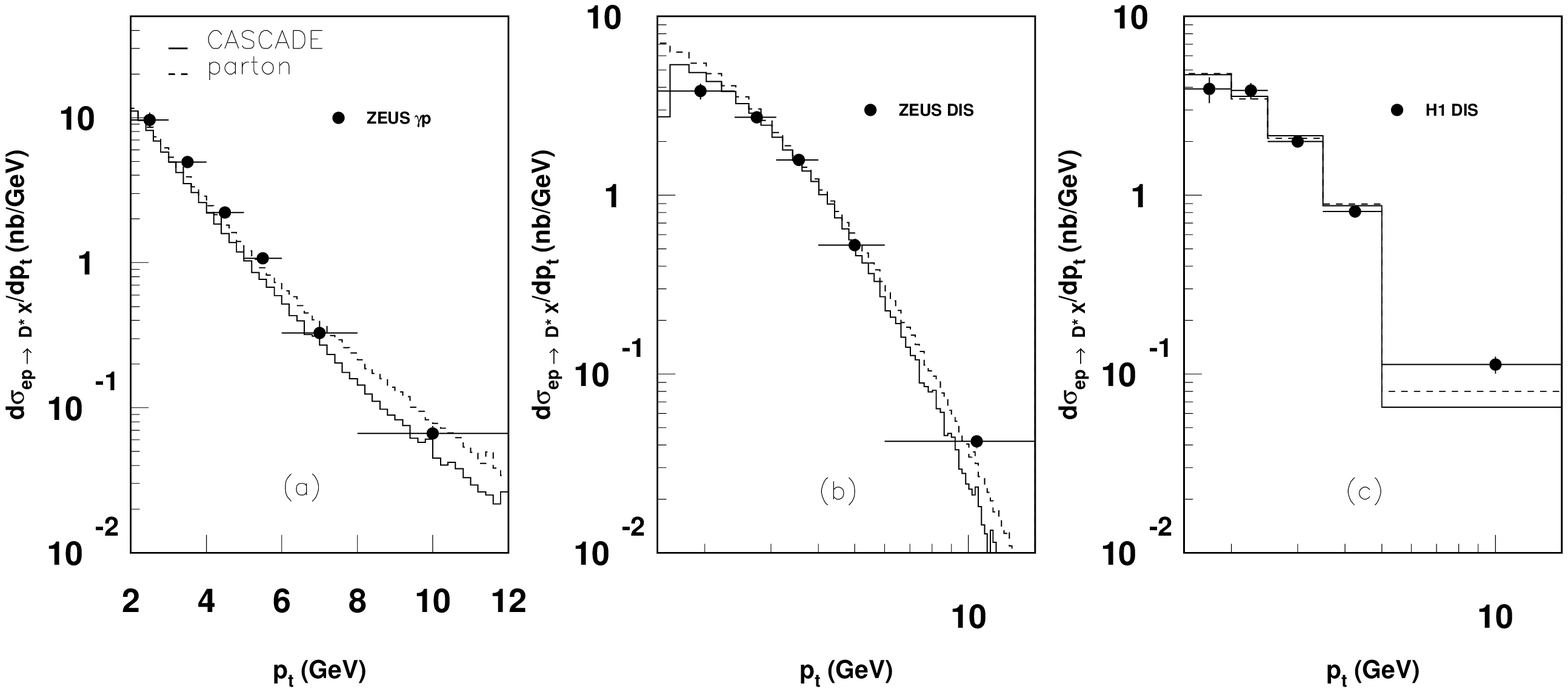,width=18cm,height=10cm}
\caption{{\it The differential cross section $d\sigma/dp_t$ for 
$D^*$ production: 
$(a)$ in photo-production (ZEUS~\protect\cite{ZEUS-D*-gammap}), 
$(b)$ in DIS (ZEUS~\protect\cite{ZEUS-D*-dis}) and 
$(c)$ in DIS (H1~\protect\cite{H1-D*-dis}).
The solid line is the prediction from the full hadron level simulation 
 \CASCADE\  and the dashed line shows the
parton level calculation. In both cases the Peterson fragmentation function has
been used.
}}\label{pt_dstar}
\end{center}
\vskip -0.5cm
\end{figure}
In Fig.~\ref{pt_dstar} we show the transverse momentum distribution of $D^*$
mesons as measured by the ZEUS~\cite{ZEUS-D*-gammap,ZEUS-D*-dis}
and H1~\cite{H1-D*-dis} collaborations both in photo-production  
and deep-inelastic scattering. The data are compared to the predictions 
of the \CASCADE\ Monte Carlo event generator on hadron level including 
a full simulation of the partonic and hadronic final state. Also shown is the
pure parton level calculation using the matrix element of 
BZ.
In both cases  the transition from the charm quark to the observed 
$D^*$ meson was performed by a simple  Peterson fragmentation function
(with $\epsilon=0.06$ and a $c \to D^*$ branching ratio  $BR=0.26$). The
scale $\bar{\mu}^2$ in $\alpha_s(\bar{\mu}^2)$ was set to 
$\bar{\mu}^2 = p_t^2 + m_c^2$ with $p_t$ being
the transverse momentum in the $\gamma g$ c.m.s.
of the final charm quark state assuming $m_c=1.5$~GeV.
The {\it JS} unintegrated gluon distribution~\cite{jung_salam_2000} was used, 
with the scale $\mu$ 
(being related to the maximum angle) $\mu^2 \sim x_{\gamma} x_g s$.
The sensitivity to the details of the charm fragmentation and to the full
initial state gluon cascade simulation can be seen by comparing \CASCADE\ with
the parton level calculation. We observe, 
that the $p_t$ distribution of $D^*$
mesons both in photo-production and deep inelastic scattering is in general well
described, both with the full hadron level simulation as
implemented in \CASCADE\ and also
with the parton level calculation supplemented with the Peterson
fragmentation function. We can thus conclude, that the $p_t$ distribution is 
only little sensitive to the details of the charm fragmentation. 
\subsection{Transverse momentum distribution of {\boldmath$D^*$} mesons: 
sensitivity to unintegrated gluon distributions}
\begin{figure}[htb]
\vskip -0.5cm
\begin{center}
\epsfig{figure=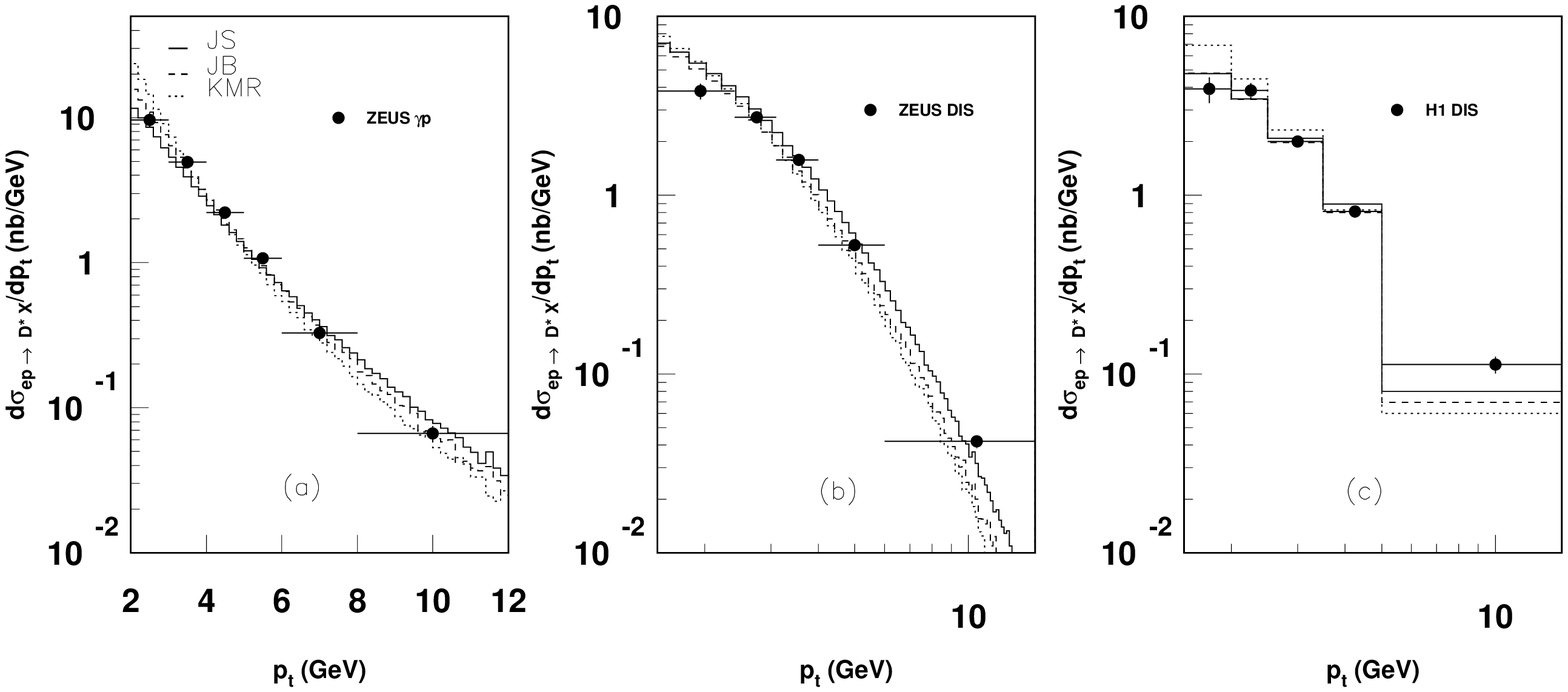,width=18cm,height=10cm}
\caption{{\it The differential cross section $d\sigma/dp_t$ for 
$D^*$ production: 
$(a)$ in photo-production (ZEUS~\protect\cite{ZEUS-D*-gammap}), 
$(b)$ in DIS (ZEUS~\protect\cite{ZEUS-D*-dis}) and 
$(c)$ in DIS (H1~\protect\cite{H1-D*-dis}).
The solid (dashed, dotted) line 
corresponds to using the {\it JS (JB, KMR)} unintegrated gluon density
(all calculated at parton level).
 In all cases the Peterson fragmentation function has
been used.
}}\label{pt_dstar_ccfm_bfkl}
\end{center}
\vskip -0.5cm
\end{figure}
Since the transverse momentum distribution of $D^*$ mesons is only slightly
sensitive to the details of the full parton cascade and charm fragmentation, 
we can now proceed to investigate the sensitivity to   
the choice of the unintegrated gluon distribution.  
In Fig.~\ref{pt_dstar_ccfm_bfkl} we show the prediction for $d\sigma/dp_t$ 
obtained from the parton level calculation as above using    
the {\it JS}, {\it JB} and {\it KMR} unintegrated gluon distributions 
in comparison with the data.  
Although differences are observed in the $x_g$ and $k_t$ distributions between  
the different unintegrated gluon distributions (see Fig.~\ref{gluon}),  
it is interesting to note, that 
very similar predictions for the $D^*$ cross
sections as a function of the transverse momentum $p_t$ are obtained.
\par
\begin{figure}[htb]
\begin{center}
\vspace*{2mm}
\epsfig{figure=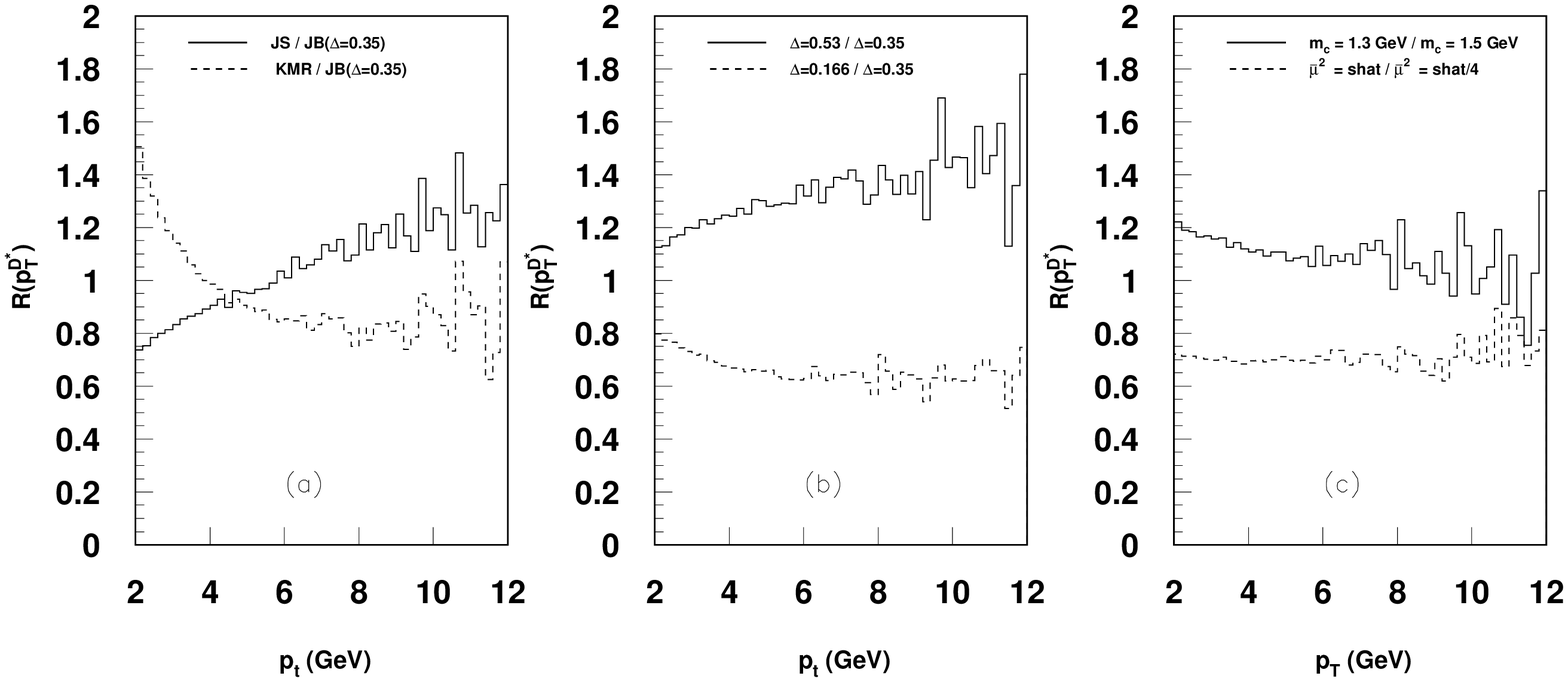,width=18cm,height=10cm}
\caption{{\it The ratio 
$R=\frac{d\sigma /dp_t }{d\sigma^{ref} /dp_t}$
as a function of $p_t$ for 
$D^*$ photo-production. The reference set corresponds to the {\it JB} set with
$\Delta= 0.35$, $m_c = 1.5$~GeV and $\bar{\mu}^2 = \hat{s}/4$.
In $(a)$ the solid (dashed) line corresponds to the {\it JS (KMR)}
unintegrated gluon distribution. In $(b)$ the parameter $\Delta$ of the {\it JB}
unintegrated gluon distribution is varied (the solid (dashed) line 
corresponds to $\Delta = 0.53$ $(0.166)$). In $(c)$ the solid line corresponds to
$m_c=1.3$~GeV and the dashed line to the setting $\bar{\mu}^2=\hat{s}$.
}\label{dstarpt_bfkl}}
\end{center}
\end{figure}

In Fig.~\ref{dstarpt_bfkl} we investigate in more detail the 
different unintegrated gluon distributions,  the effect of 
varying the $\Delta$ parameter (see sec.(3)) 
in the {\it JB} distribution and 
of changing the charm-quark mass and the evolution scale $\bar{\mu}$.
The comparison is performed on the parton level and the Peterson
fragmentation function has been used to produce the $D^*$ meson.
We define the ratio
$$R=\frac{d\sigma /dp_t }{d\sigma^{ref} /dp_t}$$
where $d\sigma^{ref} /dp_t$ 
is calculated using $\Delta=0.35$, 
$m_c=1.5$~GeV and $\bar{\mu}^2 = \hat{s}/4$.
\par
The ratio of the differential cross section as a function of $p_t$ 
obtained from the parton level calculation supplemented with the Peterson
fragmentation function using the {\it JS} (solid line) and 
{\it KMR} (dashed line)
unintegrated gluon distributions is shown in Fig.~\ref{dstarpt_bfkl}$a$. 
The fact that the ratios show different behavior is directly connected to the
different slopes in the $k_t$ distribution of the parton densities (see
Fig.~\ref{gluon}).
\par
In Fig.~\ref{dstarpt_bfkl}$b$ we show the ratio $R$ for different values of
the $\Delta$ parameter of the {\it JB} unintegrated gluon distribution. We
observe that the ratio $R$ varies with $p_t$ in the low
$p_t$-range but seems to flatten off at higher $p_t$. For $\Delta$-values larger
than the reference value  an increase in $R$ is observed, whereas 
$\Delta$-values below the reference value result in a decreasing $R$. 
This implies that
the $p_t$ spectra get harder with increasing $\Delta$-values and softer
with decreasing $\Delta$ values compared to the reference value 
$\Delta=0.35$~\cite{d*gam}.
\par
In Fig.~\ref{dstarpt_bfkl}$c$ the effects of changing the mass of 
the charm quark and the
evolution scale are illustrated. A decrease of the charm quark mass
leads to an increase of the ratio $R$ in the low $p_t$ range whereas an
increase of the interaction scale from $\bar{\mu}^2=\hat{s}/4$ to 
$\bar{\mu}^2=\hat{s}$ 
leads to an overall decrease of $R$ by $20-25$ $\%$.
\subsection{Inclusive distribution: 
comparison of parton and hadron level}
\begin{figure}[htb]
\vspace*{2mm}
\begin{center}
\epsfig{figure=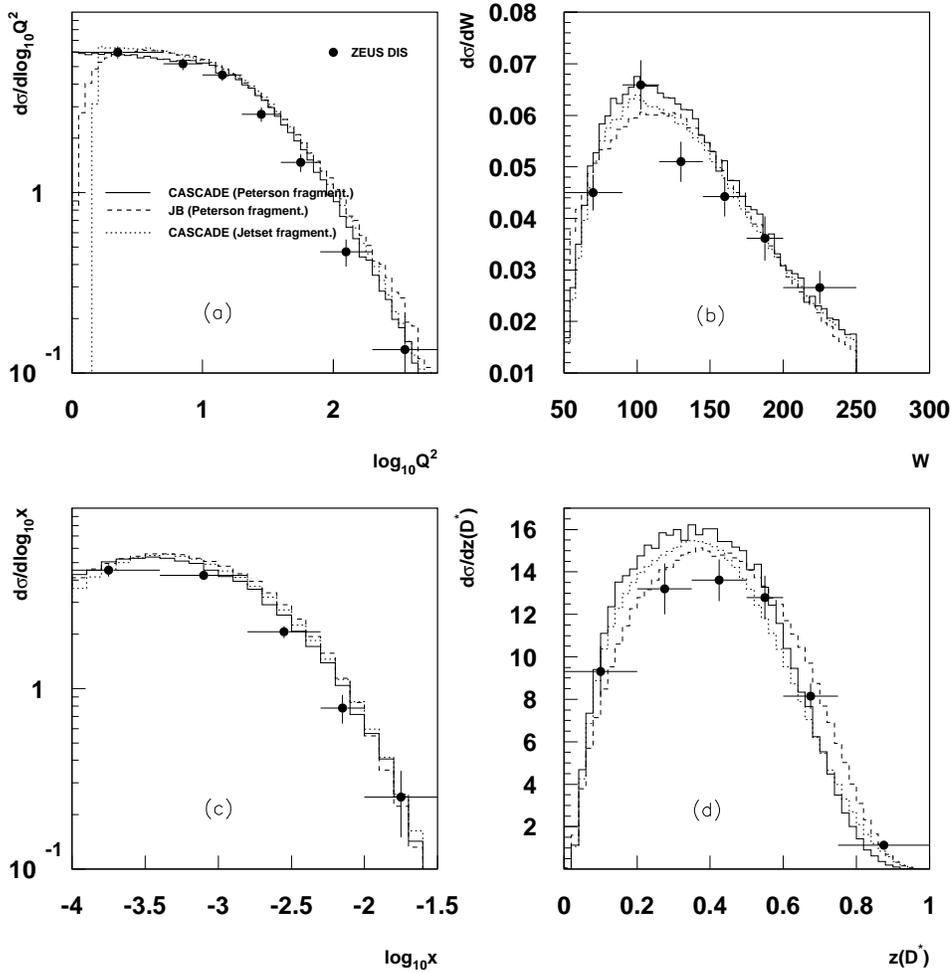,width=15cm,height=15cm}
\caption{{\it Differential cross sections  as measured by ZEUS.
Shown is a comparison of the calculations using \CASCADE\  with the
{\it JS} (solid line) and {\it JB} (dashed line) unintegrated gluon density. The
dotted line shows the \CASCADE\  prediction using the \JETSET\ charm
fragmentation function.
}}\label{zeus_cascade-bfkl}
\end{center} 
\end{figure}

In Fig.~\ref{zeus_cascade-bfkl} we compare the measured cross section
for $D^*$-production as function of $Q^2$, $W$, $x_{Bj}$ and $z_D$ with
calculations using \CASCADE\ with the {\it JS} unintegrated gluon 
distribution
(solid line) and the parton level calculation supplemented with the Peterson
fragmentation function using the {\it JB} unintegrated gluon density 
(dashed line). We also show the effect of changing the 
charm fragmentation (dotted line).
\par
Good agreement with data is observed for both \CASCADE\ with the {\it JS}
unintegrated gluon density as well as  for 
the parton level calculation with the {\it JB} 
($\Delta$=0.35) unintegrated gluon density
in the differential cross sections as a function of $\log Q^2$ and 
$\log x_{Bj}$~\cite{d*dis2}. 
The differential cross section as a function of $W$, as 
shown in Fig.~\ref{zeus_cascade-bfkl}$b$, is well
described by \CASCADE\  and somewhat less well described by {\it JB}
in the peak region, although the errors of the measurement are
fairly large. 
For the energy fraction $z_D$ taken by the $D^*$
meson, presented in Fig.~\ref{zeus_cascade-bfkl}$d$,
we observe a slight shift of the
{\it JB} distribution towards higher $z_D$ values compared to
\CASCADE . However, the 
$z_D$ distribution is sensitive to the details of the $D^*$
fragmentation, which is indicated by the dotted line in 
Fig.~\ref{zeus_cascade-bfkl}$d$, which represents \CASCADE , but using
the \JETSET\ fragmentation function instead of the Peterson one.
In conclusion of the above comparisons, we observe a good description of the
differential cross sections as a function of the transverse momentum $p_t$ of
the $D^*$ meson as well as of inclusive quantities: the unintegrated gluon
distributions, which have been considered,
 are reasonable for the description of the data. 
\subsection{Rapidity distribution of {\boldmath $D^*$} mesons: 
comparison of parton and hadron level}
\begin{figure}[htb]
\vspace*{2mm}
\epsfig{figure=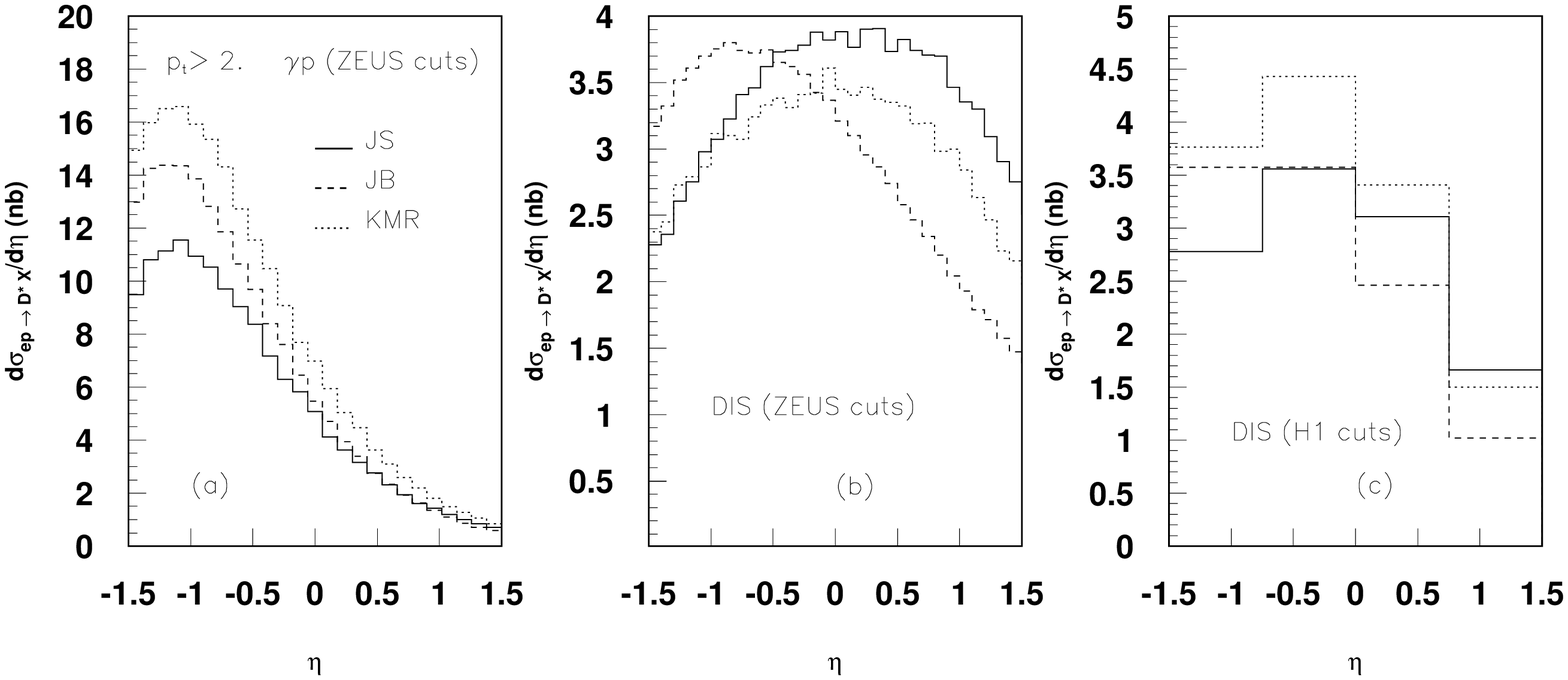,width=18cm,height=10cm}
\caption{{\it The differential cross section $d\sigma/d \eta$ for $D^*$
production: 
$(a)$ in photo-production (ZEUS cuts), 
$(b)$ in DIS (ZEUS cuts) and 
$(c)$ in DIS (H1 cuts).
Shown is a comparison of the calculations using the
{\it JS} (solid line), {\it JB} (dashed line)
and  {\it KMR} (dotted line) unintegrated gluon density.
}}\label{dstar_eta_comp_ccfm-bfkl}
\end{figure}
In photo-production and in DIS the differential cross section 
$d \sigma /d \eta$, where $\eta$ is the pseudo-rapidity of the $D^*$ meson, is
sensitive to the choice of the unintegrated gluon distribution. 
In Fig.~\ref{dstar_eta_comp_ccfm-bfkl} we show a comparison of 
$d \sigma /d \eta$ in $\gamma p$ and in DIS at parton level supplemented 
with the Peterson fragmentation function using the {\it JS} (solid line),  
{\it JB} (dashed line) and the {\it KMR} (dotted line)
unintegrated gluon distribution. Large differences in $d \sigma /d \eta$ are
visible, but one has to 
keep in mind that especially the $\eta$ distribution is also sensitive
to the details of the $c\to D^*$ fragmentation, and therefore a clear
distinction of the unintegrated gluon distributions based on this quantity alone 
might be questionable. 
\begin{figure}[htb]
\vspace*{2mm}
\epsfig{figure=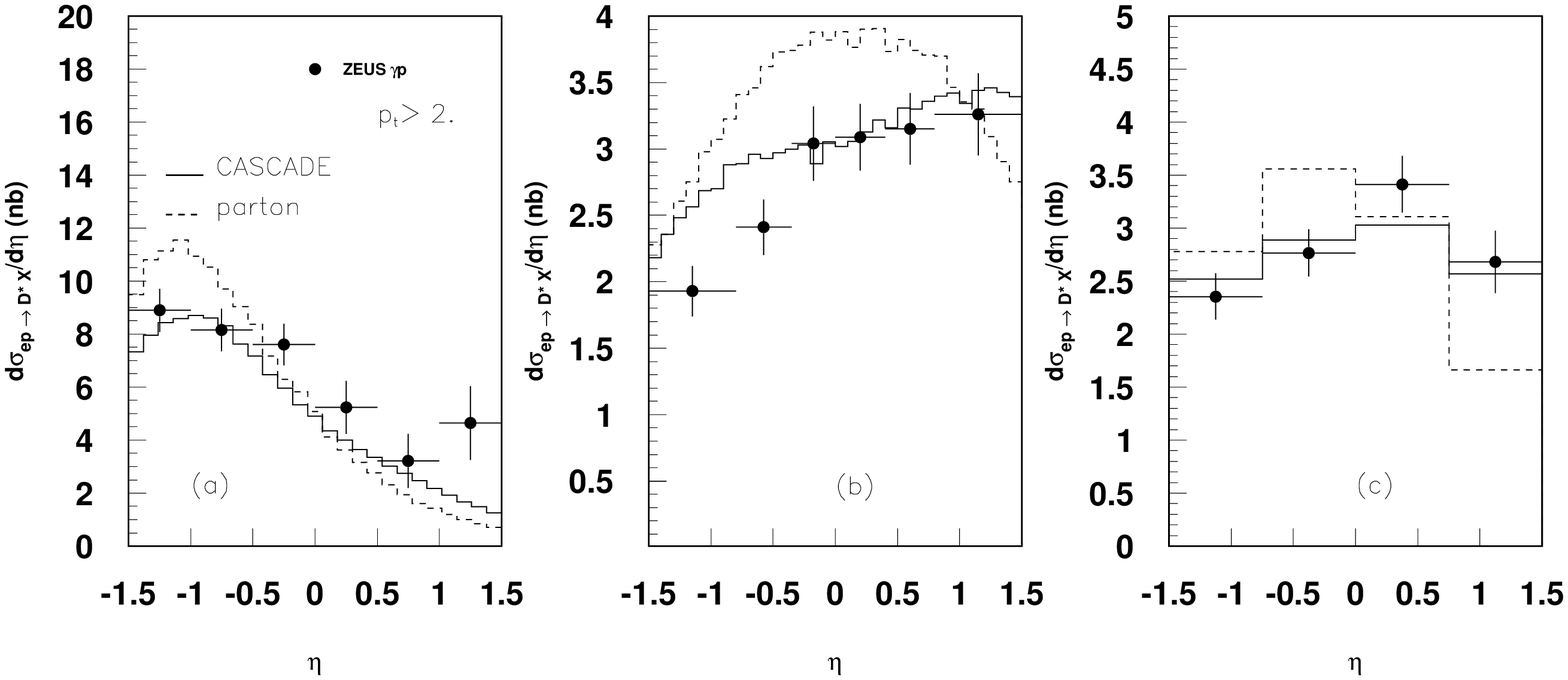,width=18cm,height=10cm}
\caption{{\it The differential cross section $d\sigma/d \eta$ for $D^*$
for  
$(a)$ photoproduction as measured by ZEUS~\protect\cite{ZEUS-D*-gammap},  
$(b)$ DIS from ZEUS~\protect\cite{ZEUS-D*-dis} and  
$(c)$ DIS from H1 ~\protect\cite{H1-D*-dis}. The histograms 
show the full hadron level simulation from CASCADE (solid line)
compared to the parton level calculation (dashed line), both
using the {\it JS} unintegrated gluon density .
}}\label{dstar_eta_comp}
\end{figure}
In Fig.~\ref{dstar_eta_comp} we show $d \sigma /d \eta$, in photo-production
and in DIS, using the {\it JS} unintegrated gluon
distribution at parton level and with the full simulation of \CASCADE .
We observe, that the parton level
prediction including the Peterson fragmentation function is not able to describe
the measurements over the full range of $\eta$. The effect of a full hadron
level simulation is clearly visible as \CASCADE\ provides a much better
description of the experimental data. 
Here the \JETSET\ charm fragmentation has been used.
\par
From the  above it
is obvious, that the $\eta$ distribution is sensitive to the details of the
fragmentation of the charm quark into the $D^*$ meson but also sensitive to the
simulation of the initial state QCD cascade. It also shows, that an
ordinary next-to-leading calculation at parton level cannot be expected to
describe the $\eta$ distribution, since a full simulation of 
the initial state parton evolution obviously is important for a 
reasonable description of heavy quark decays.
\begin{figure}[htb]
\vspace*{2mm}
\begin{center}
\epsfig{figure=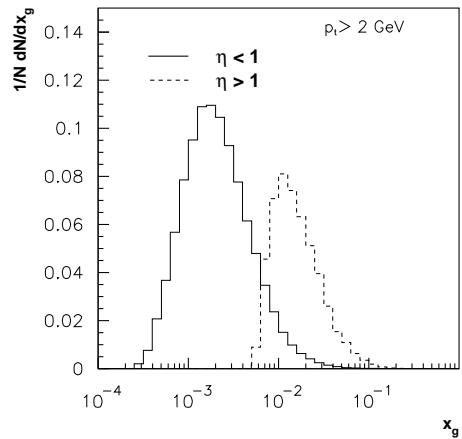,width=8cm,height=8cm}
\caption{{\it The distribution of $x_g$ in $D^*$ photo-production for $\eta  
<1$ and $\eta >1 $ for $p_{t\;D^*}>2$ GeV.
}}\label{dstar_eta-gluon}
\end{center}
\end{figure}
However, in photoproduction even the full event
simulation of \CASCADE\ shows differences to the measurement at large
values of $\eta > 1$. 
These large values of $\eta > 1 $ are related to large values of
the momentum fraction $x_g \gsim 0.03$ of the gluons entering 
the hard subprocess, as shown in Fig.~\ref{dstar_eta-gluon}. It is intuitively
understandable, that the small-$x$ approximation becomes less reliable at large
$x_g$-values. In addition, values of $x_g \gsim 0.03$ are
not constrained in the determination of {\it JS} as described in
\cite{jung_salam_2000,jung-hq-2001}. It is therefore not surprising that
the description is not perfect in this kinematic region.
\par
\begin{figure}[htb]
\vspace*{2mm}
\epsfig{figure=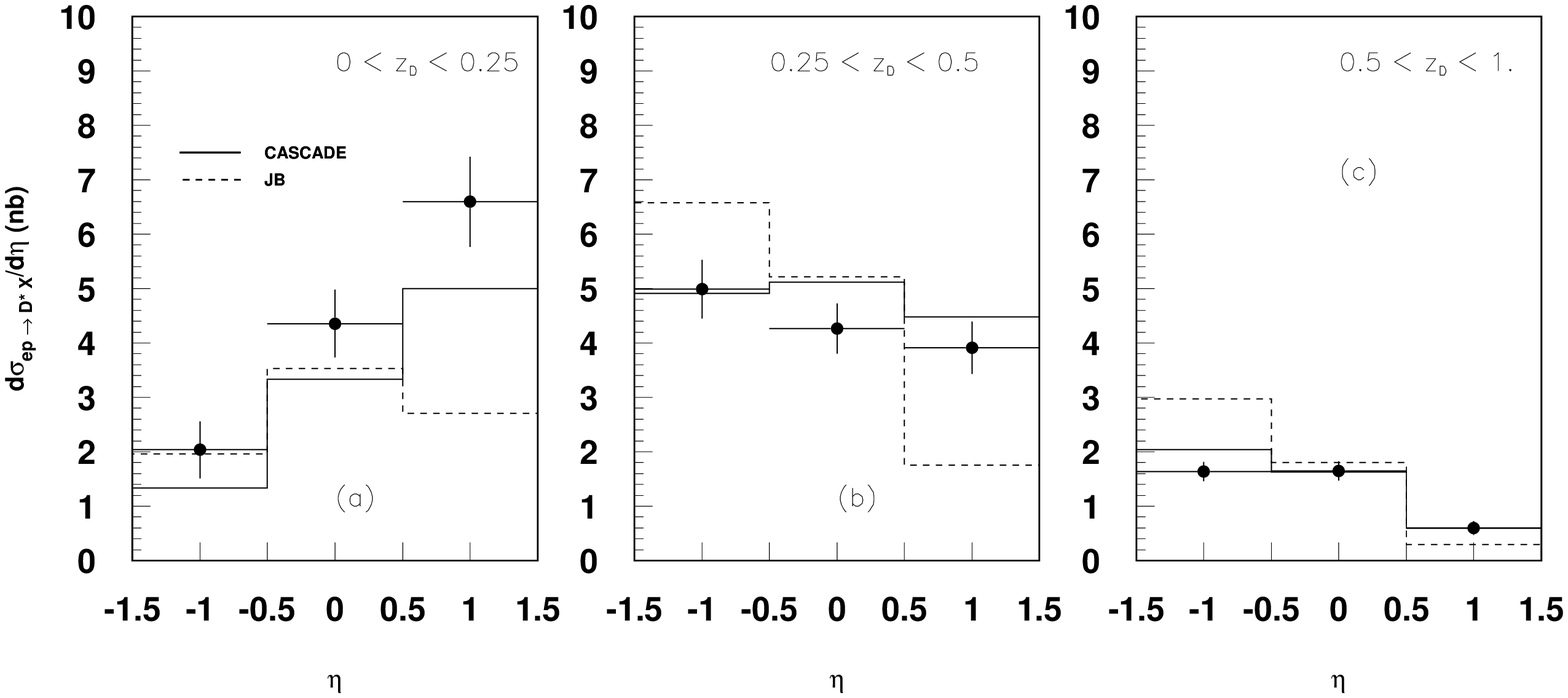,width=18cm,height=10cm}
\caption{{\it The differential cross section $d\sigma/d \eta$
for $D^*$ production in three bins
of $z_D$ as measured by H1~\protect\cite{H1-D*-dis}.
The data are compared to 
the calculations using \CASCADE\ 
with the \JETSET\ charm fragmentation and with the 
{\it JS} (solid line) unintegrated gluon density. The dashed line shows the
parton level calculation with the Peterson charm fragmentation function and with
the {\it JB} unintegrated gluon density. 
}}\label{eta_h1_cascade-bfkl}
\end{figure}

In Fig.~\ref{eta_h1_cascade-bfkl} we show the cross section 
$d\sigma/d \eta$ of deep inelastic $D^*$ production as measured by 
H1~\cite{H1-D*-dis} for different ranges in $z_D$ together with the prediction 
of \CASCADE\ with the \JETSET\ charm fragmentation function
and the {\it JS} unintegrated gluon density (solid line).
Also shown is  the parton level
calculation supplemented with the Peterson fragmentation function   
and the {\it JB} (dashed line) unintegrated gluon density. We observe that 
again the {\it JS} unintegrated gluon distribution together with a full event
simulation gives a reasonable description of the data of this double
differential cross section. 
\subsection{{\boldmath$D^*$} and associated jet production: 
comparison of parton and hadron level}
In the BFKL and/or CCFM equations the transverse momenta of
the exchanged or emitted partons are only restricted by kinematics. 
In such a scenario, the hardest $p_t$ emission
can be anywhere in the gluon chain, and needs not to sit closest to the photon
as required by the strong $q^2$ ordering in DGLAP. 
\par
Photo-production of charm is an ideal testing ground for studying the underlying
parton dynamics, 
since charm quarks are predominantly produced via $\gamma \to c
\bar{c}$. The observation of any emission (jet) with 
$p_t > p_t^c(p_t^{\bar{c}})$ indicates a scenario, which 
in DGLAP is possible only in a
full $O(\alpha_s^2)$ calculation or when charm excitation of the photon is
included. However, in $k_t$ factorization such a scenario comes naturally, since
the transverse momenta along the evolution chain are not 
$k_t$ ordered. 
\par
The ZEUS collaboration has measured charm and associated jet 
production~\cite{ZEUS-D*-gammap}. 
In these measurements, the quantity of interest is the fractional photon 
momentum involved in the production of the two jets of highest $E_t$, 
which is experimentally defined as 
\begin{equation} 
x^{OBS}_{\gamma}=\frac{E_{1t}\exp(-\eta_1)+E_{2t}\exp(-\eta_2)}{2E_e\,y }
\label{xgam} \end{equation} 
with $E_{it}$ and $\eta_i$ being the transverse energy and rapidity of the 
hardest jets and $y$ being the fractional photon energy.  
\begin{figure}[htb]
\vspace*{2mm}
\includegraphics[width=0.95\linewidth]{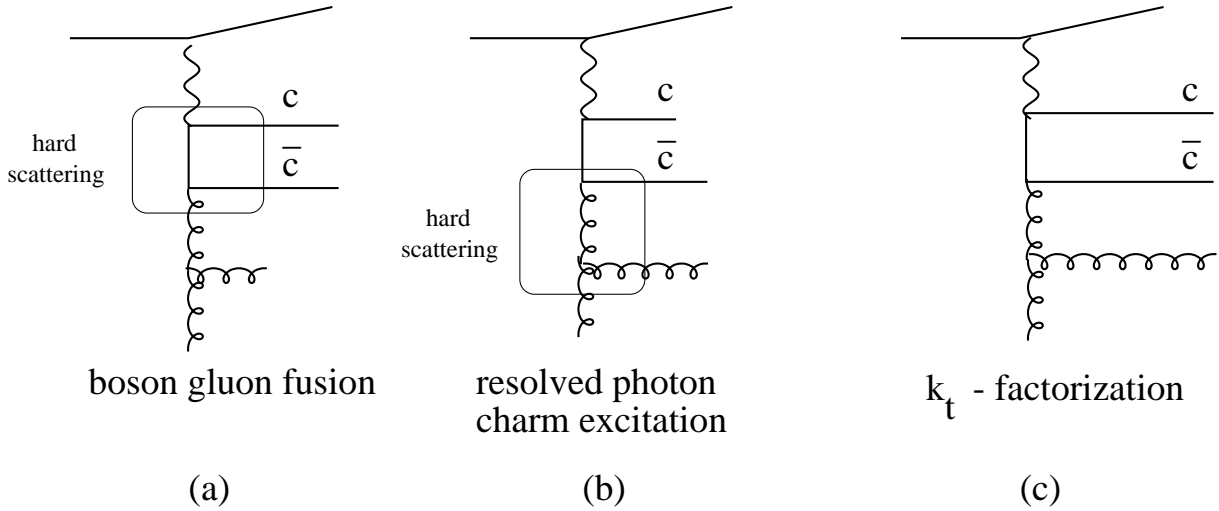}
\caption{{\it Diagrammatic representation of the processes involved in charm
photoproduction. $(a)$ shows the typical boson-gluon fusion diagram, the hard
partons are the $c \bar{c}$ quarks.
$(b)$ shows a typical resolved photon (charm excitation) diagram, the hard parton
are here the $\bar{c} g$.
$(c)$ shows a typical diagram in $k_t$-factorization, any two of $c \bar{c} g$
can be the hardest partons.
}}\label{kt-fact}
\end{figure}
In Fig.~\ref{kt-fact} we show a diagrammatic representation of the different
processes involved in charm photoproduction.
Thus, if the two hardest transverse momentum jets are produced by the $c
\bar{c}$ pair, then $x^{OBS}_{\gamma}$ is close to unity
 (Fig.~\ref{kt-fact}$a$),
but if a gluon from the initial state
cascade together with one of the $c$-quarks form the hardest 
transverse momentum
jets, then $x^{OBS}_{\gamma}<1 $ (Fig.~\ref{kt-fact}$b$). 
In a leading order calculation using the collinear
approximation $x^{OBS}_{\gamma}<1 $ indicates a resolved photon like process
(Fig.~\ref{kt-fact}$b$). 
Such a scenario is
obtained naturally also in a full NLO (${\cal O}(\alpha_s^2)$) calculation,
because in the three parton final state ($c \bar{c} g$) any of these
 partons are allowed
to take any kinematically accessible value (Fig.~\ref{kt-fact}$c$).
 In the $k_t$ factorization 
approach the anomalous component of the photon ($\gamma \to c\bar{c}$) is
automatically included, since there is no restriction on 
the transverse momenta along the evolution chain. 
\par
\begin{figure}[htb]
\begin{center}
\epsfig{figure=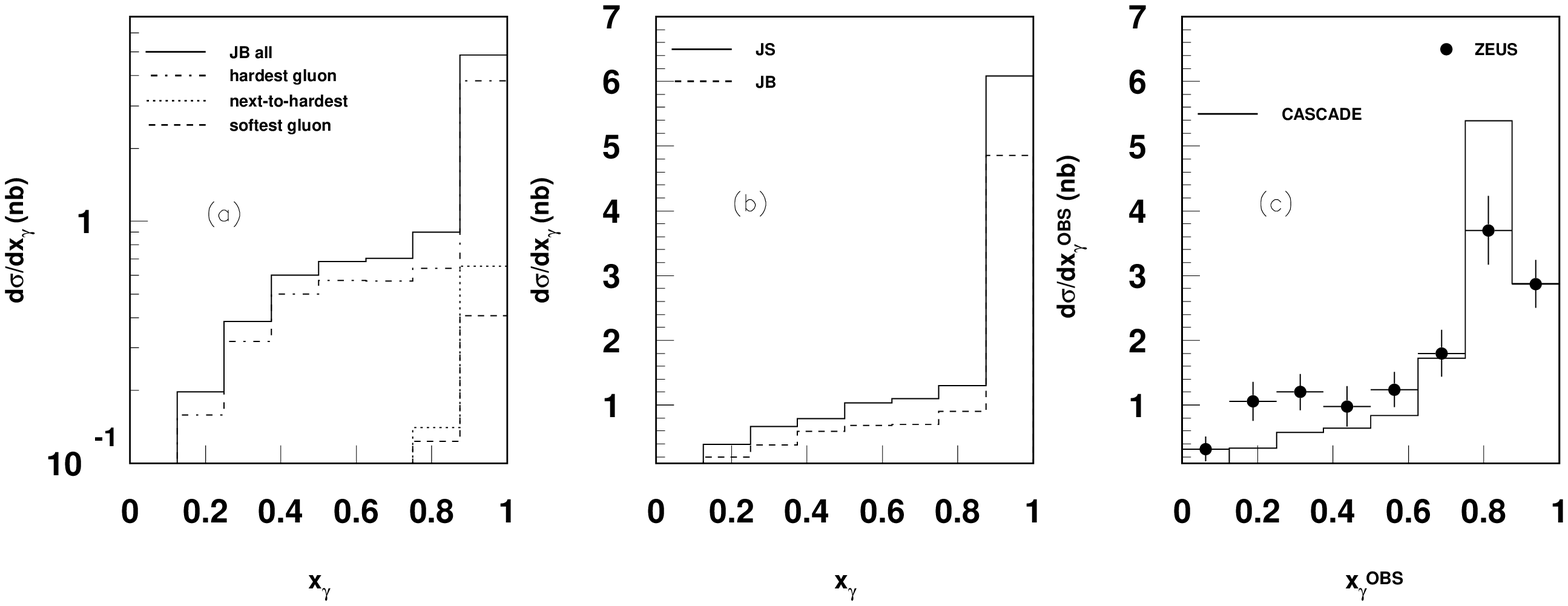,width=19cm,height=9cm}
\end{center}
\vspace*{-5mm}
\caption{{\it The differential cross section $d \sigma /dx_{\gamma}$ for 
$Q^2<1$ GeV$^2$.
In $(a)$ the parton level calculation is shown using the {\it JB} unintegrated
gluon density (solid line). Shown are also the contributions where the 
gluon takes the
largest (dash-dotted line), next-to-largest (dashed line) and 
smallest (dotted line) transverse momentum. 
In $(b)$ the {\it JB} (solid line) and {\it JS} (dashed line)
 sets are used in the parton level calculation.
In $(c)$ the measurement of ZEUS~\protect\cite{ZEUS-D*-gammap} 
is compared to the full event simulation obtained from
\CASCADE\ using the {\it JS} unintegrated gluon distribution.
}}
 \label{xgam_fig} 
\end{figure}

The experimentally observed $x^{OBS}_{\gamma}$ spectrum 
(Fig.~\ref{xgam_fig}$c$) shows a tail to
small values of $x^{OBS}_{\gamma}$, indicating that the hardest emission is
indeed not always coming from the charm quarks.
In Fig.~\ref{xgam_fig}$a$ we show a comparison of the $x^{OBS}_{\gamma}$ 
distribution
obtained on parton level with the {\it JB} unintegrated gluon density. 
Indicated is
also the contribution from events where the gluon is the hardest, 
next-to-hardest and softest parton. A significant part of the cross section
comes from events, where the gluon is the jet with the largest transverse
momentum~\cite{xgam}. 
In Fig.~\ref{xgam_fig}$b$ we compare the $x^{OBS}_{\gamma}$
distribution obtained at parton level
from the {\it JB} unintegrated gluon with the one 
from {\it JS}. In both cases a significant tail towards small $x_{\gamma}$
values is observed, however the details depend on the unintegrated gluon
distribution.
In Fig.~\ref{xgam_fig}$c$ we show a comparison of the measurement 
from the ZEUS collaboration~\cite{ZEUS-D*-gammap} with
the prediction from the full event simulation of
\CASCADE\ using the {\it JS} unintegrated gluon distribution and applying 
jet reconstruction and jet selection at the hadron level.
We observe a reasonably good agreement, showing indeed  a large tail towards 
small $x^{OBS}_{\gamma}$ values in agreement with the observation from data.
 We can
conclude that the $k_t$-factorization approach effectively simulates heavy quark
excitation and indeed the hardest $p_t$ emission comes frequently from a
gluon in the initial state gluon cascade. 
\begin{figure}[htb]
\begin{center}
\epsfig{figure=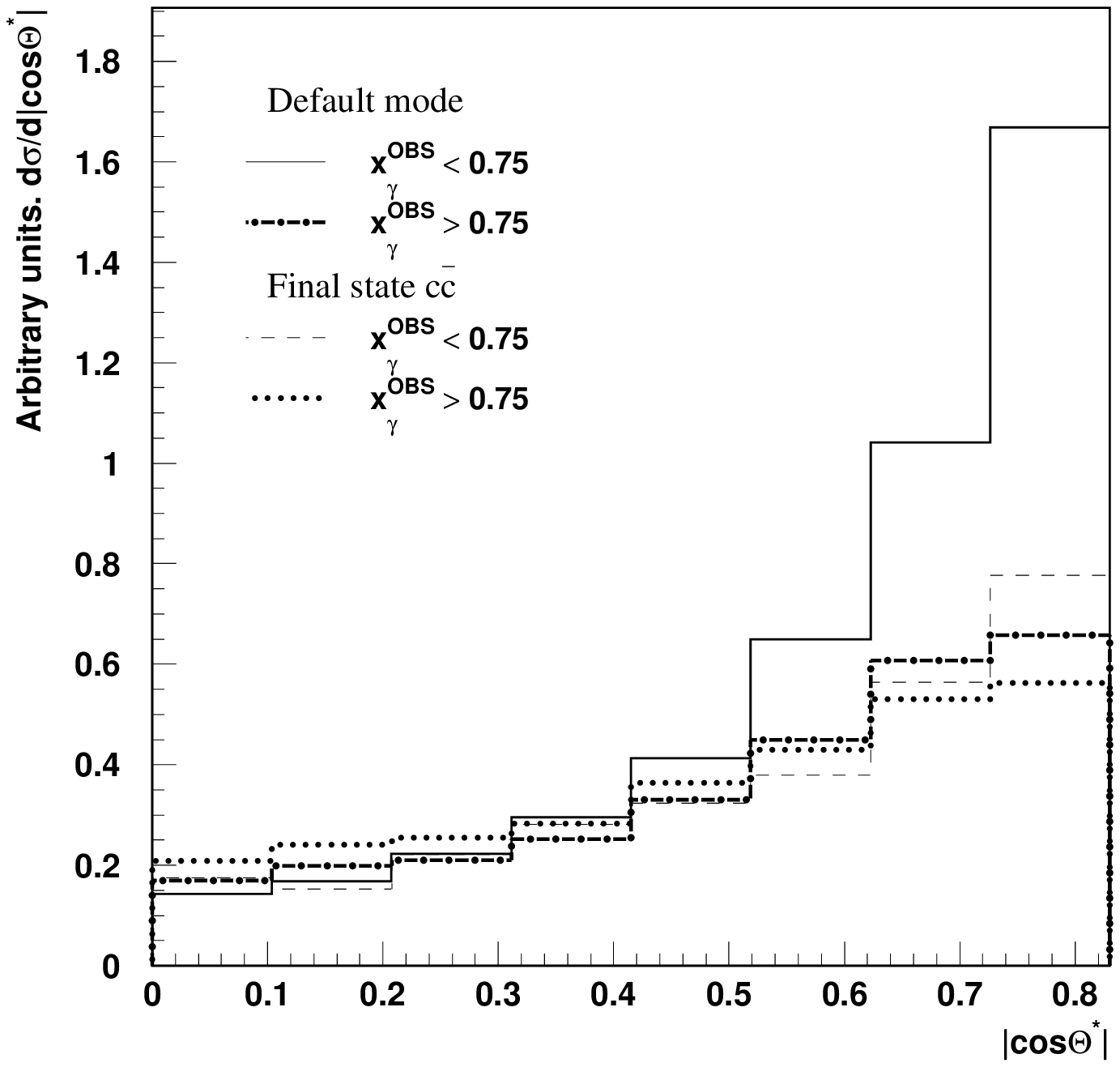,width=10cm,height=10cm}
\end{center}
\vspace*{-5mm}
\caption{{\it The distribution in $|\cos\theta^{*}|$ for
 resolved-photon like events ($x_{\gamma}^{obs} < 0.75$) and for 
direct-photon like events ($x_{\gamma}^{obs} > 0.75$)
within the kinematic range of the ZEUS  
measurement~\protect\cite{ZEUS_eps499}. \label{res_gamma_zeus}}}
\end{figure} 
\par
Another interesting quantity is the angular distribution of resolved
photon like events ($x^{OBS}_{\gamma} < 0.75$)
compared to the direct-photon  like
events ($x^{OBS}_{\gamma} > 0.75$)~\cite{ZEUS_eps499,padhi}. 
An important difference between the two (direct (Fig.~\ref{kt-fact}$a$),
resolved (Fig.~\ref{kt-fact}$b$)) 
scattering processes is that a quark (charm in this
case) is the propagator of the hard scattering 
in the direct photon like events and a
gluon in the dominant resolved photon like events. 
The angular distribution of dijets with a $D^{*\pm}$ in the final state is
dependent on the type of propagator (quark or gluon) connecting
both jets (if we neglect the case, where a parton is emitted in the rapidity 
range between the two hardest jets). In the collinear approximation, the angular
distribution is determined by the matrix element ($\gamma g \to c \bar{c}$ in
the direct case or $c g \to c g$ in the resolved photon case).
In the $k_t$-factorization approach the angular distribution will be determined
from the off-shell matrix element, which covers both scattering processes. 
\par
In  Fig.~\ref{res_gamma_zeus}   we show the $|\cos\theta^{*}|$ distribution, 
where $\theta^{*}$ is the scattering angle of the hard jets  
to the beam axis in the dijet c.m.s. Applying the same cuts and 
using the same jet-algorithm as  
in~\cite{ZEUS_eps499,padhi}, we can see that the direct photon like events
give a slow rise in cross-section with increasing
$|\cos\theta^{*}|$. However,  the cross section 
of the resolved
photon like events rises very rapidly because of the $t$-channel
gluon exchange being a combined effect of the off-shell gluon and the 
unintegrated gluon distribution. 
\par
Thus the partons of the 
initial state radiation in the 
$k_t$-factorization approach give information on  the spins of the 
propagators, as well as on the parton dynamics of the underlying
sub-processes. 
We specifically checked the dynamics in our approach
by only allowing  the final state partons ($c\bar{c}$) to
appear in the hard scattering (i.e. turning off the simulation of the initial
state gluon cascade in \CASCADE , but keeping the final state parton
shower). 
Here we should expect to see only the quark
exchange kind of behavior for both direct and resolved like events,
which is also verified as seen 
in Fig.~\ref{res_gamma_zeus} with dotted and dashed
lines overlapping around the default mode with 
$x_{\gamma} > 0.75$. 
\begin{figure}[htb]
\begin{center}
\epsfig{figure=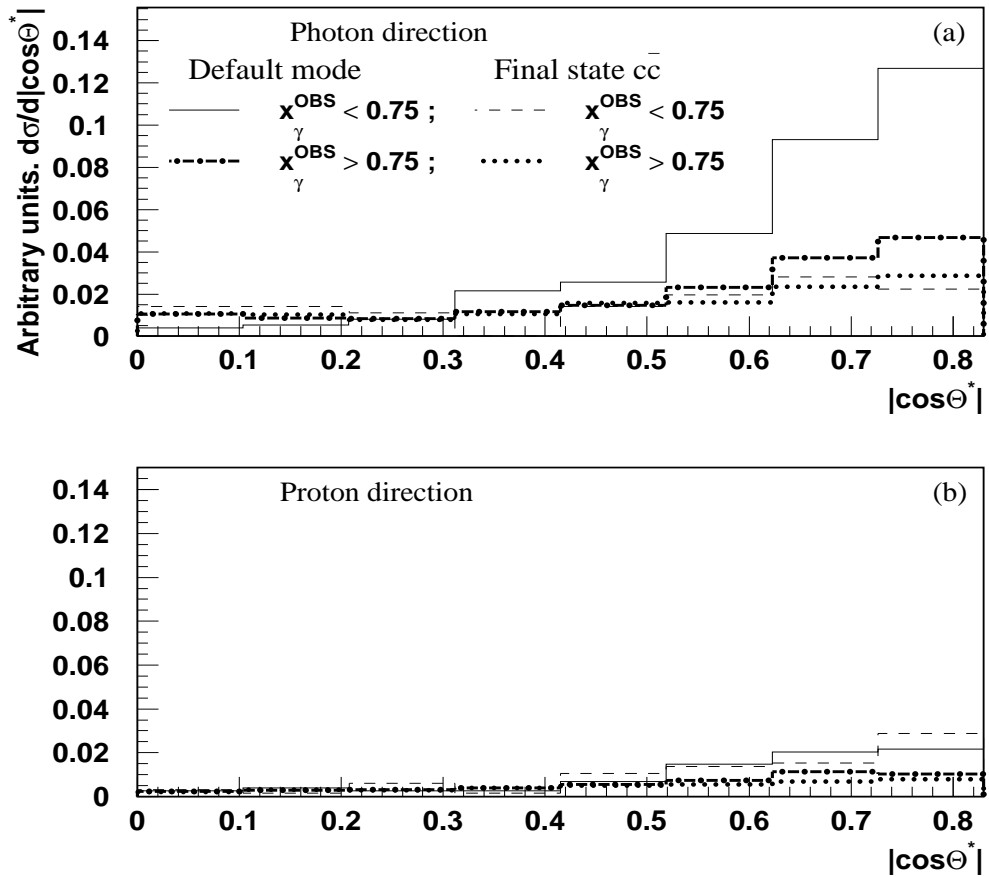,height=14cm,width=15cm}
\vspace{-0.1cm}
\caption {\it{ $|\cos\theta^{*}|$ distribution with $\eta_{D^*} < 0$ ($a$)
    (photon direction) and $\eta_{D^*} > 0$ ($b$) (proton direction) within the
    kinematic range of 
    the ZEUS measurement~\protect\cite{ZEUS_eps499,padhi} }}
    \label{fig2}
\end{center}
\end{figure}
\par
In order to further probe the parton dynamics in this approach,
we divide the entire sample into two parts, one 
where the $D^*$ proceeds along the photon
direction, (i.e. $\eta_{D^*} < 0$) and one where it travels along the 
proton direction (i.e. $\eta_{D^*}> 0$). 
If the $t$-channel gluon is indeed the dominant contribution to the cross
section, then 
the angular distribution will be
peaked towards the direction of the incoming photon. 
This asymmetry should persist (more or less) equally for charm and
anti-charm. 
Such an asymmetry cannot be seen in the inclusive dijet sample
(without separation of gluon and quark jets), because by
definition the distribution 
must be $t  \leftrightarrow  u$ symmetric as long as we
do not attempt to discriminate different kinds of jets.  
\par 
In Fig.~\ref{fig2} we show the $|\cos\theta^{*}|$ distribution
for the two cases with the $D^*$ found in the
the photon $(a)$ and proton $(b)$ direction, with $x^{OBS}_{\gamma} > 0.75$ and
$x^{OBS}_{\gamma} < 0.75$ in the same phase space as 
in~\cite{ZEUS_eps499,padhi}, except with an additional cut on average
pseudorapidity of the jets, $|\overline{\eta}|  < 0.1$, to ensure an
unbiased phase space region as discussed in \cite{padhi-phd}.
 As one can see, with the $D^*$ in the photon direction,
 there is a steep rise in the
cross-section for resolved photon like 
events compared to the direct photon like events.  
This increase in $|\cos\theta^{*}|$ obtained  
through the initial state gluon cascade in the
$k_t$-factorization approach can also be interpreted as
``charm excitation'' processes.
On the other hand, with the $D^*$ in the proton direction,
 we can only see a mild rise in cross-section for both direct
and resolved photon like events, which shows that  
the quark exchange is the dominant contribution (in the HERA kinematic range).
\section{Conclusion}
Three different models based on small $x$ resummation 
(BFKL and CCFM formalism) and
$k_t$-factorization have been studied in various aspects.
Two of the approaches only deliver results on parton level
({\it JB} and {\it KMR}) whereas the third one ({\it JS}) has been
implemented into an event generator providing complete
simulation of the initial and final state parton shower and hadronization.
The ability of the models to reproduce experimental data has
been investigated for charm production in the kinematic range
of HERA. The aim has been to find out whether optimal
sets of model parameters could be found leading to satisfactory
description of all data, but also to illustrate the sensitivity
of the model prediction from variations of the various parameters.
The unintegrated gluon distributions
of the three approaches exhibit different behaviors as a function
of $x$ but especially as a function of $k_t$. 
In spite of this we  
find good agreement between the models and data on $p_t$, $\log Q^2$,
$\log W$ and $\log x_{Bj}$ distributions, which indicates that 
these variables are not sensitive enough to differentiate between the models.
The less good description of the 
$z_D$ distribution observed for two of the models might be due to different
treatment of the $D^*$ fragmentation. The sensitivity to the $D^*$
fragmentation is seen very clearly in the case of the $\eta$-distribution
which also needs a full simulation of the initial and final 
state parton emission
to give a reasonable description of the data.
\par
From comparisons with data on photoproduction of charm leading to
jets, it became evident that a gluon from the initial cascade
frequently produces the jet of highest $p_t$.
Considering the uncertainties due to parton radiation and hadronization
effects the models give consistent results.
The distribution in polar angle of the hard jets, as generated by
the \CASCADE~program, predicts that the
gluon propagator is dominant in the hard scattering of resolved
photon like events, leading to high $p_t$-jets initiated by a quark
and a gluon. The polar angle spectrum for direct photon events,
however, is consistent with the propagator of the hard scattering
being a quark, resulting in hard quark-antiquark jets.
\par
We have shown, that the $k_t$ - factorization approach can be consistently used
to describe measurements of charm production at HERA, which are 
known to be not well reproduced
in the collinear approach. We have also shown, that in
$k_t$-factorization, resolved photon like processes are effectively simulated
including the proper angular distributions. The $k_t$ - factorization 
approach has become now
a challenging tool to understand the underlying dynamical processes in
high energy collisions.

\section*{Acknowledgments}
We are grateful to M. Derrick for careful reading of the manuscript.
S.B. and N.Z. thank the Royal Swedish Academy of Science for support.
One of us (S.P.)
acknowledges the support by Natural Sciences and Engineering
Research Council of Canada (NSERC) and thanks M. Drees, U. Karshon and
M. Wing for valuable discussions.
H.J. wants to thank the DESY directorate for
hospitality and support.

\raggedright

\end{document}